\documentclass[twocolumn,pre,superscriptaddress]{revtex4}

\usepackage{dcolumn}
\usepackage{amsmath}

\usepackage{graphicx}
\usepackage{calrsfs}

\newcommand{\etal}{{\it{}et~al.}}
\newcommand{\e}{\mathrm{e}}
\newcommand{\set}[1]{\left\lbrace#1\right\rbrace}

\begin{document}

\title{Structure of online dating markets in US cities}
\author{Elizabeth E. Bruch}
\affiliation{Department of Sociology, University of Michigan, Ann Arbor, Michigan, USA}
\affiliation{Center for the Study of Complex Systems, University of Michigan, Ann Arbor, Michigan, USA}
\affiliation{Santa Fe Institute, 1399 Hyde Park Road, Santa Fe, NM 87501, USA}
\author{M. E. J. Newman}
\affiliation{Center for the Study of Complex Systems, University of Michigan, Ann Arbor, Michigan, USA}
\affiliation{Santa Fe Institute, 1399 Hyde Park Road, Santa Fe, NM 87501, USA}
\affiliation{Department of Physics, University of Michigan, Ann Arbor, Michigan, USA}

\begin{abstract}
We study the structure of heterosexual dating markets in the United States through an analysis of the interactions of several million users of a large online dating web site, applying recently developed network analysis methods to the pattern of messages exchanged among users. Our analysis shows that the strongest driver of romantic interaction at the national level is simple geographic proximity, but at the local level other demographic factors come into play. We find that dating markets in each city are partitioned into submarkets along lines of age and ethnicity. Sex ratio varies widely between submarkets, with younger submarkets having more men and fewer women than older ones. There is also a noticeable tendency for minorities, especially women, to be younger than the average in older submarkets, and our analysis reveals how this kind of racial stratification arises through the messaging decisions of both men and women. Our study illustrates how network techniques applied to online interactions can reveal the aggregate effects of individual behavior on social structure.
\end{abstract}

\maketitle

\section{Introduction}
\label{sec:intro}
Patterns of romantic and sexual partnerships---who pairs with whom---have broad implications for health and society.  For example, the level of assortative mating (the extent to which like pairs with like) has long been considered an indicator of societal openness~\cite{glass1954, kalmijn1991}.  Mating patterns also determine how wealth and resources are passed from one generation to another, and hence persistence or change in inequality over time~\cite{schwartz2010, breen2013}, have implications for mental and physical health~\cite{waldron1996, smith2008}, and shape the sexual networks that drive the spread of sexually transmitted infections~\cite{morris1995,Liljeros01}.

There exists an extensive empirical and theoretical literature exploring the mechanisms behind patterns of romantic pairing~\cite{kalmijn1998,schwartz2013}. In societies where people choose their own mates, it is widely accepted that romantic pairing is driven by the interplay between individuals' preferences for partners and the composition of the pool of potential mates~\cite{becker1973, mare1991,xie2015}.  The  process can be modeled game theoretically as a market in which individuals aim to find the best match they can subject to the preferences of others~\cite{gale1962, roth1992}. There is also a large body of empirical work that documents the relationship between observed partnering patterns and the supply of partners, as reflected in the population composition of cities, regions, or countries~\cite{blau1982, angrist2002,lichter1995,lichter1991, blossfeld2003,guzzo2006,harknett2008,trent2011,south1992,south1992a}.

These studies are limited, however, in what they can reveal about the structure of dating or marriage markets. One issue is that we typically do not have access to the actual population of available dating partners and must instead make do with proxies such as census data, obliging us to treat entire towns or cities as a single undifferentiated market.  A more fundamental problem is that previous studies have only looked at extant partnerships, and not the larger set of all courtship interactions among mate-seeking individuals.  In order to properly study dating markets, one needs data on all courtship overtures that occur within the focal population, not only those that are successful and result in a partnership but also those that are rejected.  As we show in this paper, the complete set of such overtures forms a connected network whose structure can be analyzed to reveal key features of romantic markets. 

Unfortunately, complete data on courtship interactions have been historically hard to come by because unrequited overtures are rarely documented.  The few empirical studies that have directly observed courtship patterns have tended to focus narrowly on specific institutions, subpopulations, or geographic locations~\cite{laumann2004,sprecher1984}, and relatively little is known about the empirical structure of romantic and sexual markets across the general population or how this structure varies from one locale to another. 

The advent of online dating and its spectacular rise in popularity over the last two decades has, however, created a new opportunity to study courtship behaviors in unprecedented detail~\cite{rosenfeld2011}.  Here we report on a quantitative study of the structure of adult romantic relationship markets in the United States using nationwide data on online dating users and their behaviors.  We combine activity data for millions of participants with recently developed network analysis methods to shed light on the features of relationship markets at the largest scales.  There have been recent studies using early-stage patterns of online mate choice---who browses, contacts, or responds to whom---to shed light on individuals' preferences for mates~\cite{HHA10,Lin2013,lewis2013,bruch2016}, but the work presented here goes beyond these studies to examine how individuals' choices aggregate collectively to create structured relationship markets that strongly influence individuals' dating experiences.

\section{Results}
\label{sec:results}
The data we analyze come from a popular US online dating web site with over 4 million active users at the time of our study.  The data are described in detail in Section~\ref{sec:data} and Appendix~\ref{app:data}.  Our analysis focuses on all (self-identified) heterosexual, single men and women who sent or received at least one message on the site during the period of observation, January 1 to 31, 2014, and who indicated that they were pursuing some form of romantic relationship (long-term dating, short-term dating, and/or sex).  For each user we have a range of self-reported personal characteristics along with time-stamped records of all messages exchanged on the site. It is the latter that are the primary focus of our analysis, since it is the messaging patterns that reveal the aggregate demand for individuals within the market.

We quantify messaging patterns using methods of network analysis~\cite{Newman18}. We examine the set of all reciprocal interactions between opposite-sex users, meaning pairs of individuals such that at least one message was sent in each direction between the pair.  Reciprocal interactions we take to be a signal of a baseline level of mutual interest between potential dating partners. Our primary focus is on understanding the division of the online dating population into distinct submarkets: how does the market divide into subpopulations of daters and how are those subpopulations characterized?  We define submarkets as roughly self-contained groups of individuals within the network such that most reciprocal exchange of messages occurs within groups.  This corresponds closely to the established concept of ``community structure'' in network theory, a community in this context being a tightly knit subgroup of individuals within a larger network.  A number of sensitive techniques for the detection of network communities have been developed in recent years~\cite{Fortunato10}, and we employ a selection of those techniques here.  Technical details of the algorithmic methods used in our calculations are given in Section~\ref{sec:detection} and Appendix~\ref{app:analysis}.

\subsection{Dating markets are divided into distinct geographic regions}
For our first analysis, we examine community structure within the entire data set of all users of the web site during the month of observation.  A total of $15\,302\,512$ reciprocal interactions took place during this period.  We aggregate these interactions at the level of 3-digit ZIP codes---geographic regions used by the US Post Office---and count the number of interactions that take place between every pair of 3-digit ZIPs.  For instance, there were $75\,686$ reciprocal interactions between individuals in Manhattan and individuals in neighboring Brooklyn, but only 2170 interactions between individuals in Manhattan and individuals in far-away San Francisco.

\begin{figure}
\centering
\includegraphics[width=\columnwidth]{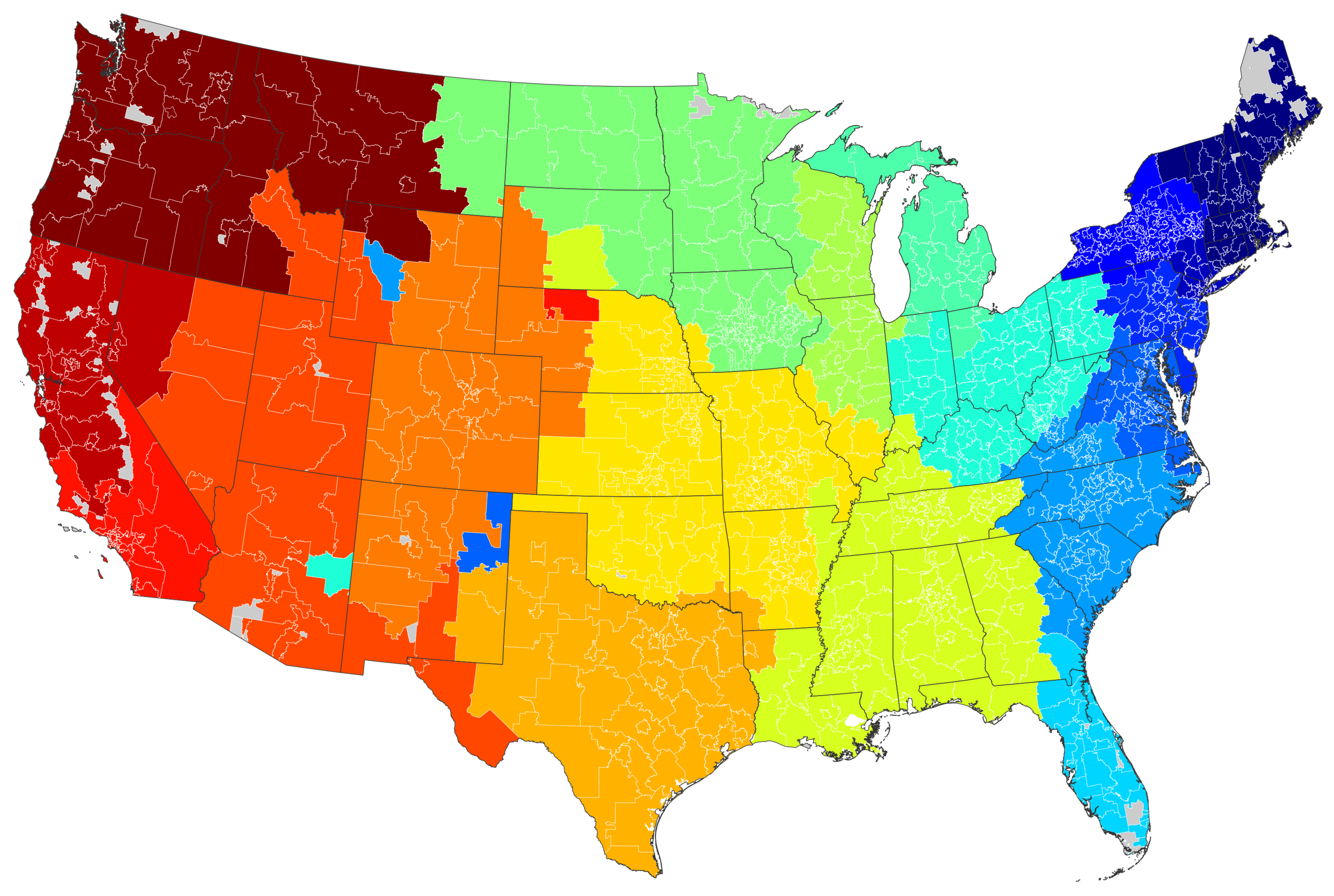}
\caption{Division of the messaging network for the lower 48 states into 19 communities by modularity maximization.  Colors represent communities at the level of 3-digit ZIP codes.  Gray regions are areas with no assigned ZIP code.}
\label{fig:communities}
\end{figure}

The result of this aggregation is a weighted network in which the nodes represent 3-digit ZIP code regions and the weighted edges represent the number of interactions.  We take this network and perform a standard community detection analysis on it using the modularity maximization method (see Section~\ref{sec:detection} and Appendix~\ref{app:analysis} for details).  The results for the lower 48 states are shown in map form in Fig.~\ref{fig:communities}.
 
As the figure shows, the communities found in this nationwide network of messaging are tightly geographically circumscribed.  Many of them appear to correspond to commonly accepted geographic divisions of the country: New England, the East Coast, the South, Texas, the Mountain West, North and South California, and so forth.  In essence the analysis says that most people are interested in others who are in their own region of the country, which is reasonable.  Few people living in New York will exchange messages with people across the country in California if the primary goal is to arrange a face-to-face meeting with a possible romantic partner~\cite{note1}.  This finding is consistent with recent work looking at friendship communities using Facebook data, which finds that incidence of friendship is strongly decreasing with geographic distance~\cite{bailey2018,note2}.

Community structure in the broad, nationwide network of messaging thus appears to be dominated by geographic effects.  Since our primary goal here is to observe and analyze more subtle demographic effects within dating markets, we need to factor out the gross influence of geography.  Our approach for doing this is a simple one: we focus on subnetworks within individual cities.  We choose cities as our basic unit of analysis because they are large enough to provide a population of significant size, yet small enough that travel distance between individuals will not be a deterrent to interaction.  In the remainder of this paper we perform a series of analyses on subsets of the data corresponding to four large cities: New York, Boston, Chicago, and Seattle.  We define cities using the standard Core-based Statistical Areas (CBSAs) for the corresponding metropolitan regions, except for New York, where the CBSA is large enough that there are clearly separate dating markets within it.  For New York we therefore define our area of study more narrowly to be the five boroughs of Manhattan, the Bronx, Queens, Brooklyn, and Staten Island.

\begin{figure*}
\centering
\includegraphics[width=10.5cm]{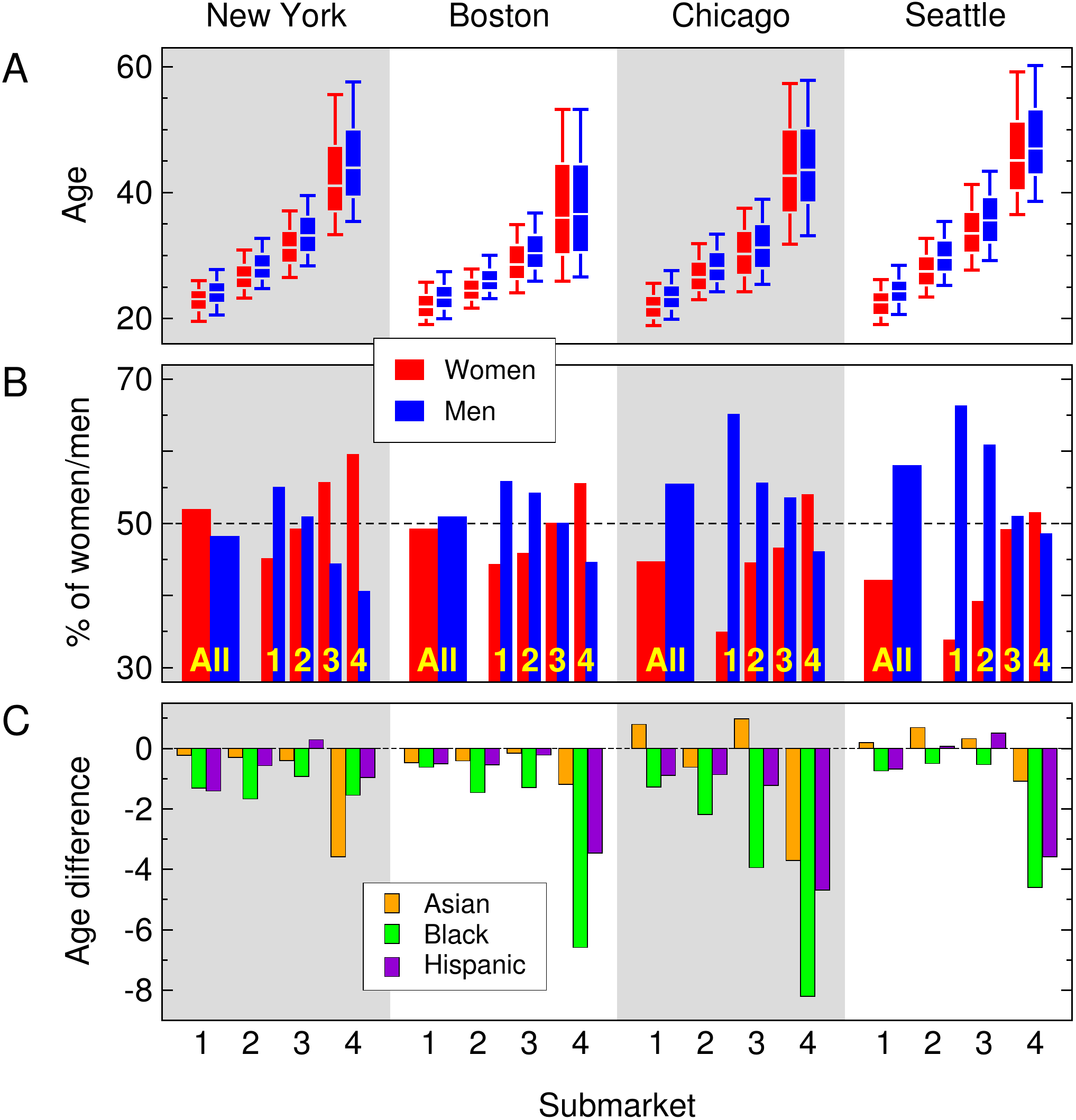}
\caption{(A)~Distribution of ages of men and women in each submarket in each of the four cities studied.  Boxes indicate 25th, 50th, and 75th percentiles; whiskers indicate 9th and 91st.  (B)~Percentage of men and women in each submarket, and overall, for each city.  (C)~Average age of minority women in each submarket by ethnicity, measured relative to average age of white women in the same submarket. Units of analysis are users.}
\label{fig:submarkets}
\end{figure*}

\subsection{Dating markets are demographically stratified within cities}
\label{sec:stratification}
Community structure at the city level is more complex than the simple geographic effects we saw in Fig.~\ref{fig:communities}.  Specifically, it displays a mix of so-called assortative and disassortative mixing~\cite{Newman03c}.  For the heterosexual dating communities studied here it is disassortative by gender, meaning most messages are between individuals of opposite sex, but assortative by various other characteristics, as we will see.  It is the latter behavior on which we primarily focus, but our community detection calculations need to be sensitive to both in order to fully reveal the structure of the market.  Here we make use of a powerful and flexible community detection method based on maximum-likelihood techniques, the expectation-maximization (EM) algorithm, and belief propagation~\cite{KN11a,DKMZ11b}, which can sensitively and rapidly detect complex forms of structure in large networks.  For details of the method see Appendix~\ref{app:analysis}.

Focusing again on networks of two-way message exchanges, we present in the following analyses the results of community divisions of each city network into four separate communities or submarkets (or eight if you count men and women separately). We find that about 75\% of all reciprocal interactions in our four cities are between individuals within the same submarket, indicating that the communities align well with the conventional definition: tightly-knit groups with most interaction going on within groups. The choice to divide into four submarkets is to some extent arbitrary.  We have repeated the analysis for other numbers of submarkets and find essentially similar patterns to those reported here---see Appendix~\ref{app:analysis}.  The choice of four submarkets offers a good compromise between resolution of finer details and adequate statistical power within submarkets.

Figure~\ref{fig:submarkets} shows a variety of demographic features of the submarkets in the four cities. The most obvious defining feature of the submarkets is the age of their members, shown in Fig.~\ref{fig:submarkets}A.  The youngest submarket, numbered~1 in each city, corresponds primarily to individuals in their lower 20s, while submarkets 2 to 4 correspond respectively to upper 20s, 30s, and 40s and above.  This pattern is consistent, with only minor variation, across the four cities.  As the figure shows, there is a small but systematic difference in age between men and women across all submarkets: in every case the men are older than the women, with a median age difference of 1 year and 7 months.

However, submarkets are not characterized by age alone.  As Fig.~\ref{fig:submarkets}B shows, they also differ in male-to-female ratio, and here we see another consistent pattern: the younger submarkets tend to be male-heavy but the mix becomes progressively more female-heavy in the older submarkets.  There are a number of factors that may drive this pattern.  Women's first marriages are at a younger age on average than men's~\cite{QP93,FP14}, which takes more women than men out of younger dating markets.  Furthermore, since partnering of younger women with older men is more common than the reverse~\cite{HBB85,HHA10}, some older men may seek out younger partners, swelling the ranks of men in the younger submarkets.  Conversely, some younger women may leave the youngest submarkets in search of older partners, depleting the supply of women.  (This would also help explain the higher average age of men in each submarket.)  The same behaviors also reduce the number of men in the older submarkets and increase the number of women.  Depending on the overall population balance of the city, the end result can be a severe distortion of the sex ratio at the oldest or youngest ages.  The youngest submarkets in Chicago and Seattle, for example, have almost two men for every woman. 

A further facet of the submarket structure, one that affects predominantly women, comes to light when we look at the balance of ethnicities.  Figure~\ref{fig:submarkets}C shows the mean age of minority women in each submarket broken down by ethnicity and measured, in this case, relative to the mean age of white women in the same submarket.  The plot demonstrates a systematic tendency for minority women to be younger than their white counterparts within the same submarket.  The effect is small in the younger submarkets but becomes more pronounced in the older ones. This is partly due to the fact that there are fewer black women than white women among the oldest users of the site (see Appendix~\ref{app:additional}, Fig.~\ref{fig:Chicago_raceComp}), but these compositional effects are not large enough to account for the pronounced age difference seen in Fig.~\ref{fig:submarkets}C.  Studies of mate preferences of online daters have shown that black women are on average viewed by heterosexual men as less desirable partners than nonblack women~\cite{fisman2008,Lin2013,HHA10,robnett2011}, and the behavior seen in Fig.~\ref{fig:submarkets}C may reflect the aggregate outcome of such preferences at the submarket level.  In Chicago's oldest submarket, for instance, black women are more than eight years younger on average than white women, suggesting that men in that submarket are exchanging messages with black women who are substantially younger than the white women they exchange messages with~\cite{note3}.

\subsection{Dating markets reflect the aggregated choices of individuals}
\begin{figure}
\centering
\includegraphics[height=\columnwidth,angle=90]{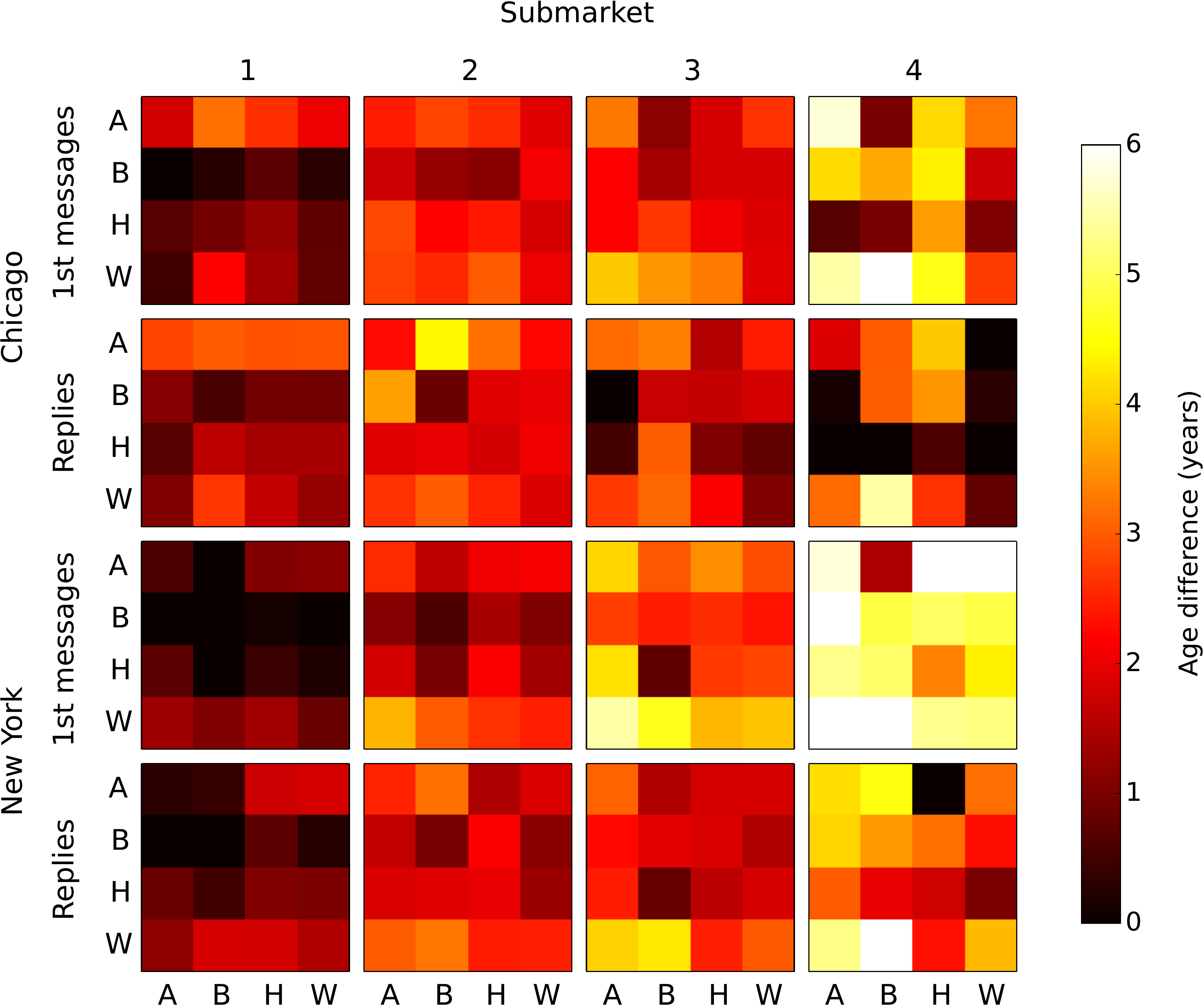}
\caption{Mean difference in years between the age of men of varying races in Chicago and New York (vertical axis) and the women they message, by race of women and by submarket (horizontal axis). Race is coded as A = Asian, B = black, H = Hispanic, and W = white.  The first two rows show the average age difference for, respectively, all initial messages sent by men in Chicago and those that received a reply.  The bottom two rows show the same patterns for New York. In both cities the age gap between men and their potential mates increases (lighter colors) as we move from younger to older submarkets. In addition, we see that black and white men in the oldest New York submarket pursue younger women, on average, than black and white men in the oldest Chicago submarket. However, unlike in Chicago, only Asian women are pursued by older black men in New York at substantially younger ages than their non-Asian counterparts. White men in the oldest submarket pursue both Asian and black women at younger ages, compared to Hispanic and white women.}
\label{fig:agediff}
\end{figure}

Next we examine how the choices of men and women about whom to message differ across submarkets, and by gender. Since men send more than 80\% of first messages on the site, we focus on men's first messages and women's replies. Figure~\ref{fig:agediff} shows the difference between the age of men and the women they message, by submarket and race, in Chicago and New York, in the form of ``heat maps.'' (Similar results for Boston and Seattle are shown in Appendix~\ref{app:additional}, Fig.~\ref{fig:BosSea_ageGap}.) The rows labeled ``1st messages'' show age difference in first messages, and the rows labeled ``Replies'' show the age difference in replies, with brighter colors corresponding to larger age differences. We see that in both Chicago and New York the age differences between men and the women they message are approximately two to three times larger in the oldest submarket than in the youngest. This is consistent with previous work showing that men's preferences for partners become more pronounced as they age~\cite{skopek2011}. 

Figure~\ref{fig:agediff} also sheds light on the behavioral mechanisms driving the racial stratification patterns we observed in Fig.~\ref{fig:submarkets}C. The top two rows of the figure for Chicago reveal that white men in older submarkets pursue minority women who are on average two or more years younger than the white women they message. This is especially pronounced in submarket~4, where the average age gap between white men and the minority women they write to is around five to six years, compared to two years for white women. However, minority women tend not to reciprocate overtures from older white men, which is why the age gap in replies among minority and white women is not as pronounced. The one exception is for black women in Chicago: the average age gap in messages between these women and the white men they respond to is around 5.8 years. Thus it is both how men pick the women they message and also how women reply that drives the racial stratification we saw in Fig.~\ref{fig:submarkets}C.

In New York the messaging patterns look somewhat different from Chicago because New York men, despite being of similar age to their Chicago counterparts, pursue younger women on average. Black men in the oldest New York submarket write to women who are on average 4.5 years younger than they are, while for white men the corresponding figure is 6.2 years. And while older white men in New York message younger black and Asian women than white women, the differences are slight: women of all races in New York's submarket~4 are being pursued at younger ages, so the racial difference is more attenuated. In other words, it's not that black women in New York's oldest submarket receive messages from younger men than black women in Chicago's oldest submarket (i.e.,~men closer to their own age), but that white women in New York's oldest submarket receive messages from older men than white women in Chicago's oldest submarket. Overall, we see that men and women's choices about who to message and respond to shape submarket structure differently in the two cities. 

\begin{figure}
\centering
\includegraphics[width=\columnwidth]{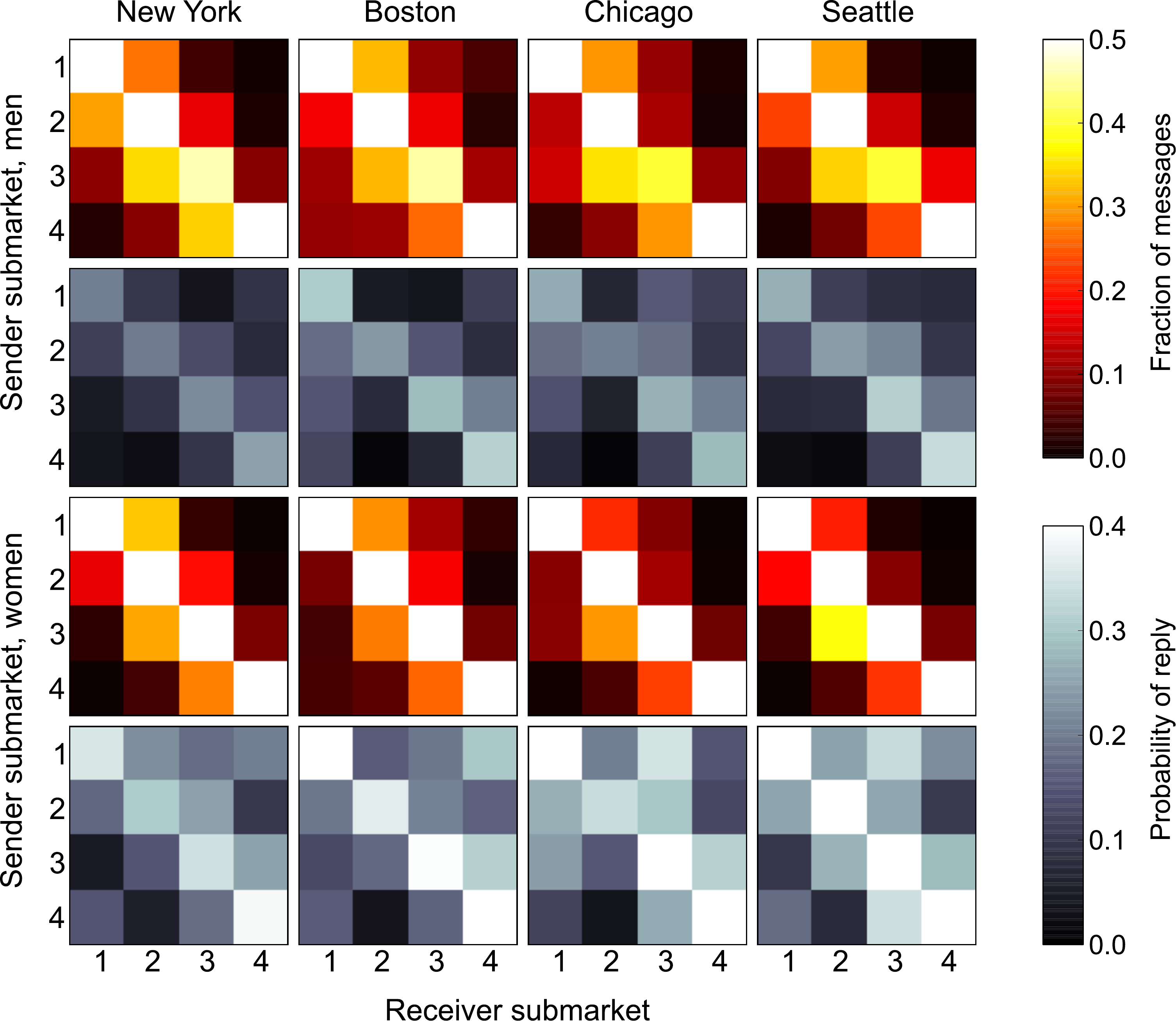}
\caption{Fraction of messages sent, and fraction that receive a reply, for messages from men to women (top two rows) and women to men (bottom two rows) in each of the four cities studied.  Brighter colors indicate larger numbers of messages.}
\label{fig:messaging}
\end{figure}

Additional features of interest in the submarket structure are revealed by an examination of messaging patterns within and between submarkets.  For this analysis we focus on initial contacts between individuals and on whether those contacts receive a reply.  Across all submarkets and cities, we find that 57\% of first contacts are between users in the same submarket.  The remaining 43\% are between users in different submarkets and the pattern of within- and between-group messages, depicted in Fig.~\ref{fig:messaging}, shows a number of interesting regularities.  The first and third rows of the figure show data for initial contacts made by men and women respectively.  The bright squares down the diagonal of each matrix represent the large fraction of within-group contacts.  The darker squares off the diagonal show that users are sending a modest number of messages to the submarkets immediately older and younger than their own, but very few messages to submarkets two or more steps away.  One deviation from this pattern is visible in the messages sent by men in submarket~3 (the 30-somethings).  Across all four of our cities, this group is the only one whose members send a majority of their messages to women in different submarkets from their own, the largest number going to women in the next youngest submarket, submarket~2 (mid-to-late 20s). 

The second and fourth rows of Fig.~\ref{fig:messaging} give the fraction of first messages that receive a reply---establishing a possible reciprocal interest between the individuals in question.  Women's replies to messages sent by men (second row) occur at a substantially lower rate than men's replies to women (fourth row), which is likely a volume effect: since women receive four times as many first messages as men, they can afford to be more selective in their replies.  Again, across all cities and among both men and women, reply rates are highest within submarkets.  Women receive replies more often when initiating contact with men in older submarkets compared to younger ones (which is consistent with prior studies), although there are some exceptions.  Notice for instance that in all cities women in the oldest submarket (submarket~4) are---surprisingly---more likely to receive a reply from men in the youngest submarket (submarket~1) than in the second youngest (submarket~2).

\section{Discussion}
The experience of mate selection is frequently described, both in popular discourse and in the scientific literature, in the language of markets: an individual's goal is to secure the best possible mate for themselves in the face of competition from others. However, we know little about the structure of these romantic markets in part for lack of appropriately detailed data.  The advent and vigorous growth of the online dating industry in the last two decades provides a new source of data about courtship interactions on an unprecedented scale. 

In this study, we have provided a first look at how network analysis techniques can reveal the structure of US dating markets as evidenced by interactions on a popular online dating website. Across the US as a whole, we find that geography is the defining feature of national dating markets. Within cities, submarkets are defined by age as well as other demographic factors---most notably, race. We find that submarket structure is shaped by both first messaging patterns and replies.  Three-quarters of all reciprocated messages fall within submarkets and only a quarter between individuals in different submarkets.  A larger fraction, about 43\%, of all first messages are between different submarkets, which indicates that people do attempt to contact partners outside of their submarkets, but that those attempts are often unsuccessful. Overall, our results reveal the aggregate implications of individuals' mate choices, and suggest that metropolitan areas are best characterized as a collection of geographically integrated but demographically distinct submarkets.

More generally, our study illustrates how state-of-the-art network science techniques can be applied to rich data from online interactions or administrative records to reveal subtle features of social structure.  In recent years the growing availability of search data from online sources has led to interest in how individuals' choices reveal submarkets in other social domains~\cite{piazzesi2015,rae2015}. As we have shown in the dating context, market outcomes reflect the choices made by actors on both sides (e.g., men and women in heterosexual dating markets, workers and firms in job markets).  Our approach could straightforwardly be extended to look at structural features of housing or job markets, and we view this as a fruitful direction for future work.

\section{Materials and methods}
\subsection{Data}
\label{sec:data}
The data used as the starting point for our study come from one of the largest free dating sites in the United States and were collected in July 2014.  The site does not market itself to any particular demographic group and attracts a diverse population of users whose makeup, in most locales, corresponds loosely to that of the general population.  The site is known for its user-driven matching algorithm, which reduces the effect of site interference on users' mate choice behavior.  The population of users is concentrated in coastal areas, although there are significant numbers of users in major Midwestern cities such as Chicago. We restrict our analysis to active users, which we define to mean that they sent or received at least one message on the site during the observation period, which was January 1 to 31, 2014.  This eliminates a significant number of users who sign up and use the site but then become inactive, or who sign up and never use it.  We also remove from the data all users who identify as gay or bisexual (about 14\% of the overall user base of the site) and those who indicate that they are not looking for romantic relationships.  (People can indicate, for example, that they are only looking for friendship or activity partners.)  Further description of the data is given in Appendix~\ref{app:data}.

\subsection{Community detection}
\label{sec:detection}
The primary technical tool employed in our analysis is community detection~\cite{Fortunato10}, which takes a network of nodes and the connections, or edges, between them---users and messages in the present context---and divides it into tightly knit groups such that most edges fall within groups and few fall between.  The most widely used method for community detection is modularity maximization~\cite{Fortunato10,Newman04a}, which makes use of the standard quality function known as modularity~\cite{NG04}.  This function, defined as the fraction of edges within groups minus the expected fraction of such edges if edges are placed at random, is large and positive for divisions of a network into good communities and small for poor divisions.  Modularity maximization finds good communities by looking for the division with the largest modularity score.  In our analysis of the complete, nationwide network of messages between active users, Fig.~\ref{fig:communities}, we make use of modularity maximization on the weighted network of conversations between users in different 3-digit ZIP codes.  There are a range of practical methods for performing the maximization itself.  In our calculations, we use the Louvain algorithm of Blondel~\etal~\cite{Blondel04}, which is an iterative greedy algorithm that has been shown to give high-quality results with short run times~\cite{YAT16}.  We use the implementation from the Gephi network analysis package, with resolution parameter equal to 0.65, which results in the 19-community division shown in Fig.~\ref{fig:communities}.

Though it is fast and gives good results, modularity maximization is not ideal for the community analysis of our individual city networks.  This is because, as discussed in Section~\ref{sec:stratification}, these networks contain both assortative and disassortative structure.  Modularity maximization is normally capable only of detecting assortative structure.  For this part of our analysis, therefore, we use an alternative community detection method based on maximum-likelihood fitting of a generative, community-structured network model, the degree-corrected stochastic block model~\cite{KN11a}.  In this approach one defines a model that generates networks with community structure, then fits that model to the observed network.  The parameters of the best fit tell us which nodes of the network belong to which communities.  More specifically they give us the posterior probability that each node belongs to each community; in the final stage of the calculation we assign every node to the community for which it has highest probability of membership.  The fitting itself is performed using an EM algorithm, with the E-step carried out using belief propagation~\cite{DKMZ11b}.  Technical details are given in Appendix~\ref{app:analysis}.  Code is available upon request from the authors.

\begin{acknowledgments}
The authors thank Travis Martin for useful conversations.  This work was funded in part by the National Institutes of Health under grant K01-HD-079554 (EEB) and the National Science Foundation under grants DMS--1107796, DMS--1407207, and DMS--1710848 (MEJN).
\end{acknowledgments}

\appendix
\section{Data}
\label{app:data}
Our data come from a popular, free online dating site. New users of the site begin by creating a profile, which includes various socio-demographic information, and they can also answer a set of open-ended essay questions that ask them to describe who they are and what they are looking for. The only information a user is required to give is their login handle, age, sexual orientation, relationship status, and a 5-digit ZIP code identifying their location. After creating a profile, users can then view the profiles of others, as well as send and receive messages. Unlike other dating sites, that are largely driven by a matching algorithm, this site allows users to pursue mates relatively freely according to their own preferences.

\subsection{Metropolitan areas}
Our city-level results are based on data from four metropolitan areas---New York, Boston, Chicago, and Seattle.  In the case of Boston, Chicago, and Seattle, we find a good choice of boundaries to be the standard Core Based Statistical Areas (CBSAs) established by the Office of Management and Budget~\cite{note4}.  For New York, however, the data clearly indicate multiple geographic dating markets within the larger metro area.  Instead, therefore, we choose a narrower set of geographic boundaries for New York, the five boroughs of Manhattan, Brooklyn, Queens, the Bronx, and Staten Island.

\begin{table*}
\label{table1}
\centering
\begin{tabular}{@{}lcccccccccc@{}}
\multicolumn{1}{c}{} &  \multicolumn{2}{c}{New York} & \multicolumn{2}{c}{Boston} & \multicolumn{2}{c}{Chicago} & \multicolumn{2}{c}{Seattle} \\
                       & Men           & Women         & Men           & Women         & Men           & Women         & Men           & Women         \\
\hline
Total number of users  & $44\,009$     & $50\,618$     & $9\,113$      & $9\,355$      & $28\,635$     & $23\,236$     & $12\,721$     & $9\,248$      \\
Ethnicity (\%)         &               &               &               &               &               &               &               &               \\
 \qquad Asian          &       8       &       11      &       4       &       6       &       3       &       4       &       7       &       9       \\
 \qquad Black          &       9       &       9       &       6       &       6       &       7       &       9       &       4       &       3       \\
 \qquad Hispanic       &       10      &       8       &       3       &       3       &       8       &       7       &       3       &       3       \\
 \qquad White          &       73      &       73      &       87      &       85      &       81      &       80      &       87      &       85      \\
College degree (\%)    &       92      &       96      &       70      &       80      &       63      &       71      &       64      &       68      \\
Children at home (\%)  &       5       &       6       &       7       &       10      &       7       &       10      &       15      &       17      \\
Mean age               &       31.6    &       31.5    &       30.4    &       30.3    &       31.4    &       32      &       32.7    &       33.1    \\
Mean messages sent     &       23.3    &       9.4     &       14.6    &       6.3     &       19      &       10.2    &       12.4    &       7.8     \\
Replies received (\%) & 15    &       34      &       17      &       37      &       18      &       40      &       20      &       45      \\
\hline
\end{tabular}
\caption{User attributes for four metropolitan areas.  Reproduced from Ref.~\cite{BruchNewman2018}.}
\label{tab:data}
\end{table*}

\subsection{Summary statistics}
Table~\ref{tab:data} provides summary statistics of users in each of the four cities, broken out by gender.  As discussed in Section~\ref{sec:stratification}, the cities vary in the ratio of men to women on the web site, New York having the largest fraction of women, followed by Boston, Chicago, and Seattle, in that order.  Recall that in Fig.~\ref{fig:submarkets}B we found the older submarkets to be more female-heavy, while the younger submarkets tended to be male-heavy.  Examination of the age distribution of men and women in each city~\cite{BruchNewman2018} suggests that this is not merely a result of age-specific sex ratios in the overall user population.  New York, for instance, has a surplus of women, which is most pronounced among younger users in their mid twenties, yet the submarkets for younger users still have significantly more men than women.  (The remaining cities all have an overall surplus of men, which is most pronounced in the later 20s and early~30s.)  These observations suggest that the submarket sex ratios observed in Fig.~\ref{fig:submarkets}B are driven by users' mate seeking behavior, and not broader population demographics.

In addition to the sex ratios, Table~\ref{tab:data} also shows that cities differ in their overall market size and composition.  New York is the largest market, followed by Chicago, Seattle, and Boston.  We also observe some variation in the average number of initial contacts made by men and women in each city, as well as their reply rates.  Consistent with other work~\cite{HHA10,Lin2013,lewis2013}, we see that men send more messages than women. However, men have a lower chance than women of receiving replies to their messages.

\section{Network analysis}
\label{app:analysis}
As described in Section~\ref{sec:results}, the starting point for our results is community structure analysis of networks of reciprocated messaging between pairs of individuals.  Our city-level analyses are restricted to the largest connected component of the network for each city, although in practice this has little effect since nearly everyone belongs to the largest component.  In the network for New York, for example, the largest connected component contains 99.8\% of all users.

Our analysis of the full, nationwide messaging network in Fig.~\ref{fig:communities} is based on standard modularity maximization, as described in Section~\ref{sec:detection}.  The structure within our individual city networks, however, is more complicated, being partly assortative (with respect to submarket) but also partly disassortative (with respect to gender, since most messages are between a man and a woman).  To correctly detect and classify this kind of mixed structure we need a more flexible detection method.  The leading such method is the statistical inference method based on fitting the network to a stochastic block model~\cite{NS01,BC09,KN11a,DKMZ11b}, which is the approach we employ in this work.  Specifically, we use the degree-corrected stochastic block model~\cite{KN11a}, which is a generative model of a random community-structured network as follows.

Let $n$ be the number of nodes in the observed network (a number typically in the thousands or tens of thousands for the networks studied here).  The degree-corrected block model allows us to create a model network of the same size by first generating $n$ nodes, numbered from 1 to~$n$, each of which is assigned to one of $k$ communities or submarkets.  The communities are numbered from 1 to~$k$, and nodes are assigned to communities independently at random, with probability~$\gamma_r$ of being assigned to community~$r$, where the $\gamma_r$ are parameters we choose, subject to the normalization constraint
\begin{equation}
\sum_{r=1}^k \gamma_r = 1.
\label{eq:sumrule}
\end{equation}
When all nodes have been assigned to communities, edges are placed at random between pairs of nodes, independently but with probabilities that depend on the communities to which the nodes belong, such that when all edges have been placed the number falling between any pair of nodes $i,j$ is Poisson distributed with mean $d_id_j\omega_{rs}$, where $r$ and $s$ are, respectively, the communities to which nodes $i$ and~$j$ belong, $\omega_{rs}$~are parameters that we choose, and $d_i$ is the degree of node~$i$ in the observed network that we are fitting (i.e.,~it is the number of connections node~$i$ has to other nodes).  The inclusion of $d_i$ is what distinguishes this ``degree-corrected'' model from other forms of the stochastic block model.  As we will see, the degree correction fixes the expected degree of every node within the model to be equal to the observed degree of the same node in the data, allowing the model to give significantly better fits to empirical data.

This defines the ``forward'' process of generating a random network given the parameters $\gamma,\omega$ of the model.  Using the model for community detection involves the inverse process of fitting the model to observed data so as to determine the values of the parameters that give the best fit.  This we do by the method of maximum likelihood.  Our undirected network of two-way communication between web site users is represented by an adjacency matrix~$A$ with elements~$a_{ij}=1$ if there is an edge between nodes $i$ and~$j$ and zero otherwise.  It is straightforward to show that the probability, or likelihood, of generating the observed network from the model, for given values of the parameters~$\gamma,\omega$, is \begin{equation} P(A|\gamma,\omega) = \sum_c P(A,c|\gamma,\omega) = \sum_c \e^{\mathcal{L}(c)},
\label{eq:likelihood}
\end{equation}
where $c$ denotes the complete set of community assignments~$\set{c_i}$ and the log-likelihood~$\mathcal{L}(c) = \log P(A,c|\gamma,\omega)$ of generating a particular set of community assignments and edges is given by
\begin{align}
\mathcal{L}(c) &= \sum_{ij} \bigl[ a_{ij} \log \omega_{c_i,c_j} - d_i d_j \omega_{c_i,c_j} \bigr] \nonumber\\
  &= \sum_{ijrs} \delta_{c_i,r} \delta_{c_j,s} \bigl[ a_{ij} \log \omega_{rs}
   - d_i d_j \omega_{rs} \bigr],
\label{eq:loglike}
\end{align}
where $\delta_{rs}$ is the Kronecker delta and we have neglected additive and multiplicative constants independent of the parameters, since they have no effect on the position of the likelihood maximum.

\subsection{Expectation-maximization (EM) algorithm}
To find the values of the parameters~$\gamma$ and~$\omega$ most likely to have generated the observed network we wish to maximize Eq.~\eqref{eq:likelihood} with respect to the parameters.  Direct maximization is cumbersome so we employ a standard trick from the machine learning toolkit.  First, we maximize not the likelihood itself but its logarithm, $\log P(A|\gamma,\omega)$, which gives the same result since the logarithm is a monotone increasing function of its argument and hence the maximum of the logarithm falls in the same place as the maximum of the argument.  Then we apply Jensen's inequality, which says that for any set of non-negative quantities~$x_i$, we have
\begin{equation}
\log \sum_i x_i \ge \sum_i q_i \log {x_i\over q_i},
\label{eq:jensen}
\end{equation}
where $q_i$ is any properly normalized probability distribution satisfying~$\sum_i q_i = 1$.  The exact equality is recovered for the special choice
\begin{equation}
q_i = {x_i\over\sum_i x_i}.
\label{eq:exact}
\end{equation}
Applying Jensen's inequality to the log of Eq.~\eqref{eq:likelihood}, we find that
\begin{align}
\log P(A|\gamma,\omega) &= \log \sum_c \e^{\mathcal{L}(c)}
   \ge \sum_c q(c) \log {\e^{\mathcal{L}(c)}\over q(c)} \nonumber\\
  &\hspace{-5em}{} = \sum_{ijrs} q^{ij}_{rs} \bigl[ a_{ij} \log \omega_{rs}
     - d_i d_j \omega_{rs} \bigr] - \sum_c q(c) \log q(c),
\label{eq:ineq}
\end{align}
where we have made use of Eq.~\eqref{eq:loglike} for the log-likelihood~$\mathcal{L}(c)$.  Here $q(c)$ is any properly-normalized probability distribution we choose over community assignments~$c$, and $q^{ij}_{rs}$ is the probability within that distribution that nodes~$i$ and~$j$ belong to communities~$r$ and~$s$ respectively, thus:
\begin{equation}
q^{ij}_{rs} = \sum_c \delta_{c_i,r} \delta_{c_j,s} \, q(c).
\label{eq:qijrs}
\end{equation}

Following Eq.~\eqref{eq:exact}, the exact equality in~\eqref{eq:ineq} is established, and hence the right-hand side maximized, when we make the choice
\begin{equation}
q(c) = {P(A,c|\gamma,\omega)\over\sum_{c'} P(A,c'|\gamma,\omega)}
     = {\e^{\mathcal{L}(c)}\over\sum_{c'}\e^{\mathcal{L}(c')}}.
\label{eq:estep}
\end{equation}
Thus if we maximize the right-hand side of~\eqref{eq:ineq} over possible choices of~$q(c)$ it becomes equal to the left-hand side, and if we further maximize the left-hand side with respect to the parameters~$\gamma,\omega$ we get the answer we are looking for---the values of $\gamma,\omega$ that maximize the overall likelihood.  Put another way, a double maximization of the right-hand side with respect to both $q(c)$ and $\omega,\gamma$ will achieve our goal.

At first sight, this appears to make the problem harder: we have turned what was previously a single maximization into a double one.  But in fact the double maximization usefully splits the problem into two parts that separately are both straightforward, whereas the original combined problem was difficult.  Maximization with respect to~$q(c)$ is achieved by making the choice~\eqref{eq:estep}, as we have said.  Maximization with respect to~$\gamma$ and $\omega$ can be achieved by simple differentiation.  Note that the final sum on the right-hand side of Eq.~\eqref{eq:ineq} does not depend on $\gamma$ or~$\omega$, so it vanishes upon differentiating.  Taking the derivative of the first sum with respect to $\gamma_r$ and~$\omega_{rs}$ while imposing the constraint~\eqref{eq:sumrule} then gives us
\begin{equation}
\gamma_r = {1\over n} \sum_i q^i_r,
\label{eq:mstepgamma}
\end{equation}
and
\begin{equation}
\omega_{rs} = {\sum_{ij} a_{ij} q^{ij}_{rs}\over\sum_i d_i q^i_r \sum_j d_j q^j_s},
\label{eq:mstepomega}
\end{equation}
where $q^i_r$ is the probability within the distribution~$q(c)$ that node~$i$ belongs to group~$r$:
\begin{equation}
q^i_r = \sum_c \delta_{c_i,r} q(c) = \sum_s q^{ij}_{rs},
\label{eq:qir}
\end{equation}
the second equality being true for any value of~$j$.

The result is an expectation-maximization or EM algorithm for fitting the model to the observed network, requiring the simultaneous solution of Eqs.~\eqref{eq:estep}, \eqref{eq:mstepgamma}, and~\eqref{eq:mstepomega}, which is accomplished by simple iteration.  We first choose initial values of the parameters~$\gamma$ and~$\omega$, for instance at random, and use them to calculate the probability distribution~$q(c)$ from Eq.~\eqref{eq:estep}.  Then we use that distribution to calculate~$q^{ij}_{rs}$ and $q^i_r$ from Eqs.~\eqref{eq:qijrs} and~\eqref{eq:qir}, and thence to calculate improved estimates of the parameters from Eqs.~\eqref{eq:mstepgamma} and~\eqref{eq:mstepomega}.  Then we recalculate $q(c)$ again, and repeat until convergence is reached.

The end product is a set of best-fit values of the parameters to the observed network data.  In addition to this, however, and crucially for our purposes, we also calculate a converged value of the distribution~$q(c)$, which, from Eq.~\eqref{eq:estep}, is equal to
\begin{equation}
q(c) = {P(A,c|\gamma,\omega)\over\sum_{c'} P(A,c'|\gamma,\omega)}
     = {P(A,c|\gamma,\omega)\over P(A|\gamma,\omega)}
     = P(c|A,\gamma,\omega).
\end{equation}
In other words, $q(c)$~is the \emph{posterior distribution} over community assignments, the probability, given the observed data~$A$ and the best-fit parameter values, of any particular division~$c$ of the network into communities.  The final step of the calculation is then to assign each node to the community for which it has the highest probability of membership, which is also equivalent to choosing the community for which $q^i_r$ is maximized.  This gives us our best division of the network into communities or submarkets.

\subsection{Expected degree}
A key feature of the degree-corrected block model is its ability to provide a good fit to networks with broad distributions of node degree (the degree of a node in a network being the number of connections it has to other nodes).  Most empirical networks, including our messaging networks, have widely varying values of node degree and any model we fit to such networks must, at a minimum, be capable of capturing this variation.

The actual degree of a node in our model network can fluctuate from one realization of the model to another, since the model contains random elements.  But the expected value of the degree of node~$i$, for the best-fit values of the parameters~$\gamma,\omega$ given in Eqs.~\eqref{eq:mstepgamma} and~\eqref{eq:mstepomega}, is always equal to the degree~$d_i$ of the same node in the observed network.  Thus the fitted network fits the degree distribution exactly apart from fluctuations.  To see this, observe that the expected degree of node~$i$ in the model is equal to the sum of the expected number of edges $d_i d_j \omega_{c_i,c_j}$ between node~$i$ and every other node~$\sum_j d_i d_j \omega_{c_i,c_j}$, averaged over the distribution~$q(c)$ of community assignments, thus:
\begin{align}
\sum_c q(c) \sum_j d_i d_j \omega_{c_i,c_j}
  &= \sum_c q(c) \sum_j d_i d_j
     \sum_{rs} \delta_{c_i,r} \delta_{c_j,s} \omega_{rs} \nonumber\\
  &= \sum_{jrs} q^{ij}_{rs} d_i d_j \omega_{rs},
\end{align}
where we have made use of Eq.~\eqref{eq:qijrs}.  Most nodes~$j$, however, will be far from node~$i$ in a large network, so that the community assignments of $i$ and $j$ are essentially uncorrelated.  This means that $q^{ij}_{rs} = q^i_r q^j_s$ and the expected degree becomes
\begin{align}
d_i & \sum_{rs} q^i_r \omega_{rs} \sum_j q^j_s d_j
   = d_i \sum_{rs} {q^i_r \sum_{ij} a_{ij} q^{ij}_{rs}\over\sum_k d_k q^k_r}
     \nonumber\\
  &= d_i \sum_r {q^i_r \sum_{ij} a_{ij} q^i_r\over\sum_k d_k q^k_r}
   = d_i \sum_r {q^i_r \sum_i d_i q^i_r\over\sum_k d_k q^k_r}
   = d_i \sum_r q^i_r \nonumber\\
  &= d_i,
\end{align}
where we have made use of Eq.~\eqref{eq:mstepomega} in the first equality, Eq.~\eqref{eq:qir} in the second, and the trivial observation $\sum_j a_{ij} = d_i$ in the third.

\subsection{Belief propagation and the calculation of the posterior distribution}
Elegant though the EM algorithm is for the community detection problem, it is not (yet) a workable method, because for all but the very smallest of networks is it not feasible to evaluate the posterior distribution~$q(c)$ directly from Eq.~\eqref{eq:estep}---the number of possible values of~$c$ is simply too large.  The number of possible divisions of $n$ nodes into~$k$ communities is $k^n$, so a division of $10\,000$ nodes into, say, four communities would have $4^{10000} \simeq 10^{6000}$ possible divisions, which is far more than can be enumerated by even the most powerful computer.  Within the statistical literature, the standard way of circumventing this problem is to approximate the distribution~$q(c)$ using Markov chain Monte Carlo importance sampling, and that could be done here too.  In our work, however, we use a recently-proposed alternative approach based on belief propagation~\cite{DKMZ11a,DKMZ11b,Yan14}, which is significantly more efficient for the particular problem at hand.

The belief propagation method focuses on a quantity~$\mu^{i\to j}_r$, called the belief, which is equal to the (posterior) probability that node~$i$ belongs to community~$r$ if we are not told whether there is an edge between nodes~$i$ and~$j$, i.e.,~if we are given the entire adjacency matrix~$A$ except for the element~$a_{ij}$.  The omission of this one matrix element is crucial to the method: it allows us to write a self-consistent set of equations for the beliefs that can be solved by numerical iteration.  For the degree-corrected block model used here, the appropriate equations have been given by Yan~\etal~\cite{Yan14}:
\begin{equation}
\mu^{i\to j}_r = {\gamma_r\over Z_{i\to j}} \exp \biggl( - \sum_k d_i d_k
  \sum_s \omega_{rs} q^k_s \biggr) \prod_{\substack{k(\ne j)\\a_{ik}=1}}
  \omega_{rs} \mu^{k\to i}_s,
\label{eq:bp1}
\end{equation}
where $Z_{i\to j}$ is a normalizing constant with value
\begin{equation}
Z_{i\to j} = \gamma_r \sum_r \exp \biggl( - \sum_k d_i d_k
  \sum_s \omega_{rs} q^k_s \biggr) \prod_{\substack{k(\ne j)\\a_{ik}=1}}
  \omega_{rs} \mu^{k\to i}_s,
\label{eq:zij}
\end{equation}
and $q^i_r$ is the one-node marginal posterior probability of node~$i$ belonging to group~$r$ defined previously in Eq.~\eqref{eq:qir}.  This probability can itself be calculated directly from the beliefs according to
\begin{equation}
q^i_r = {\gamma_r\over Z_i} \exp \biggl( - \sum_k d_i d_k
  \sum_s \omega_{rs} q^k_s \biggr) \prod_{\substack{k\\a_{ik}=1}}
  \omega_{rs} \mu^{k\to i}_s,
\label{eq:bp2}
\end{equation}
with
\begin{equation}
Z_i = \gamma_r \sum_r \exp \biggl( - \sum_k d_i d_k
  \sum_s \omega_{rs} q^k_s \biggr) \prod_{\substack{k\\a_{ik}=1}}
  \omega_{rs} \mu^{k\to i}_s.
\label{eq:zi}
\end{equation}
The belief propagation calculation involves choosing an initial set of values for the beliefs and the one-node probabilities (for instance at random in the interval~$[0,1]$), using them first to calculate new values of the~$q^i_r$ from Eqs.~\eqref{eq:bp2} and~\eqref{eq:zi}, and then using those values, plus the beliefs, to calculate new values of the beliefs from Eqs.~\eqref{eq:bp1} and~\eqref{eq:zij}.  Then we repeat the procedure, iterating until the beliefs converge.

This gives a set of beliefs for the current values of the parameters~$\gamma,\omega$.  Returning to the EM algorithm, we then use those values to compute improved estimates of the parameters from Eqs.~\eqref{eq:mstepgamma} and~\eqref{eq:mstepomega}.  To do this, we first need to calculate the two-node marginal probabilities~$q^{ij}_{rs}$ from the beliefs, which we do as follows.

Note that $q^{ij}_{rs}$ appears only in the sum in the numerator of Eq.~\eqref{eq:mstepomega} and that the sum involves only the values of~$q^{ij}_{rs}$ for node pairs $i,j$ that are connected by an edge.  (Those not connected by an edge have $a_{ij}=0$ and hence do not appear in the sum.)  For pairs connected by an edge, $q^{ij}_{rs}$~is by definition equal to
\begin{align}
q^{ij}_{rs} &= P(c_i=r, c_j=s|a_{ij}=1,A') \nonumber\\
           & = P(a_{ij}=1 | c_i=r, c_j=s,A') {P(c_i=r, c_j=s|A')\over
              P(a_{ij}=1|A')},
\end{align}
where the parameters~$\gamma,\omega$ are assumed given in each probability and $A'$ denotes the set of elements of the adjacency matrix excluding~$a_{ij}$ (which is specified separately).  But each term in this expression is now straightforward to write in terms of quantities we already know.  The probability $P(a_{ij}=1 | c_i=r, c_j=s,A')$ is just the likelihood of the edge from $i$ to $j$, which for our stochastic block model is
\begin{equation}
P(a_{ij}=1 | c_i=r, c_j=s,A') = d_i d_j \omega_{r,s} \e^{-d_i d_j \omega_{rs}}.
\end{equation}
Since $\omega_{rs}$ is typically very small, it is usually acceptable to neglect the exponential.  (Recall that we are only interested in assigning each vertex to the highest-probability community, so small errors in the probabilities typically make no difference to the final answer.)  And the probability that $c_i=r$ given~$A'$ is precisely the belief~$\mu^{i\to j}_r$, so
\begin{equation}
P(c_i=r, c_j=s|A') = \mu^{i\to j}_r \mu^{j\to i}_s.
\end{equation}
The probability $P(a_{ij}=1|A')$ is fixed by the requirement of normalization, meaning it can be calculated by stipulating that $\sum_{rs} q^{ij}_{rs} = 1$.  The end result is
\begin{equation}
q^{ij}_{rs} = {d_i d_j \omega_{rs} \mu^{i\to j}_r \mu^{j\to i}_s\over
               \sum_{rs} d_i d_j \omega_{rs} \mu^{i\to j}_r \mu^{j\to i}_s}.
\label{eq:bpq}
\end{equation}
Substituting this value into Eq.~\eqref{eq:mstepomega} now gives us our new value for~$\omega_{rs}$.

Our final, combined EM/belief propagation algorithm now consists of the following steps:
\begin{enumerate}
\item We choose initial values of the parameters~$\gamma_r$ and $\omega_{rs}$ for all $r,s$, for instance at random.
\item We choose initial values of the beliefs~$\mu^{i\to j}_r$ and one-node marginal probabilities~$q^i_r$, for instance at random.
\item We iterate the belief propagation equations, \eqref{eq:bp1}~to~\eqref{eq:zi}, to convergence to give values for the beliefs~$\mu^{i\to j}_r$ and the one-node marginal probabilities~$q^i_r$.
\item We use these values to calculate the two-node probabilities~$q^{ij}_{rs}$ from Eq.~\eqref{eq:bpq}.
\item We use the one- and two-node probabilities to calculate improved estimates of $\gamma_r$ and $\omega_{rs}$ for all $r,s$ from Eqs.~\eqref{eq:mstepgamma} and~\eqref{eq:mstepomega}.
\item We repeat steps 3 to 5 until the parameters and probabilities converge.
\item We assign each node to the community~$r$ for which its probability of membership~$q^i_r$ is highest.
\end{enumerate}

\begin{figure}
\centering
\includegraphics[width=\columnwidth]{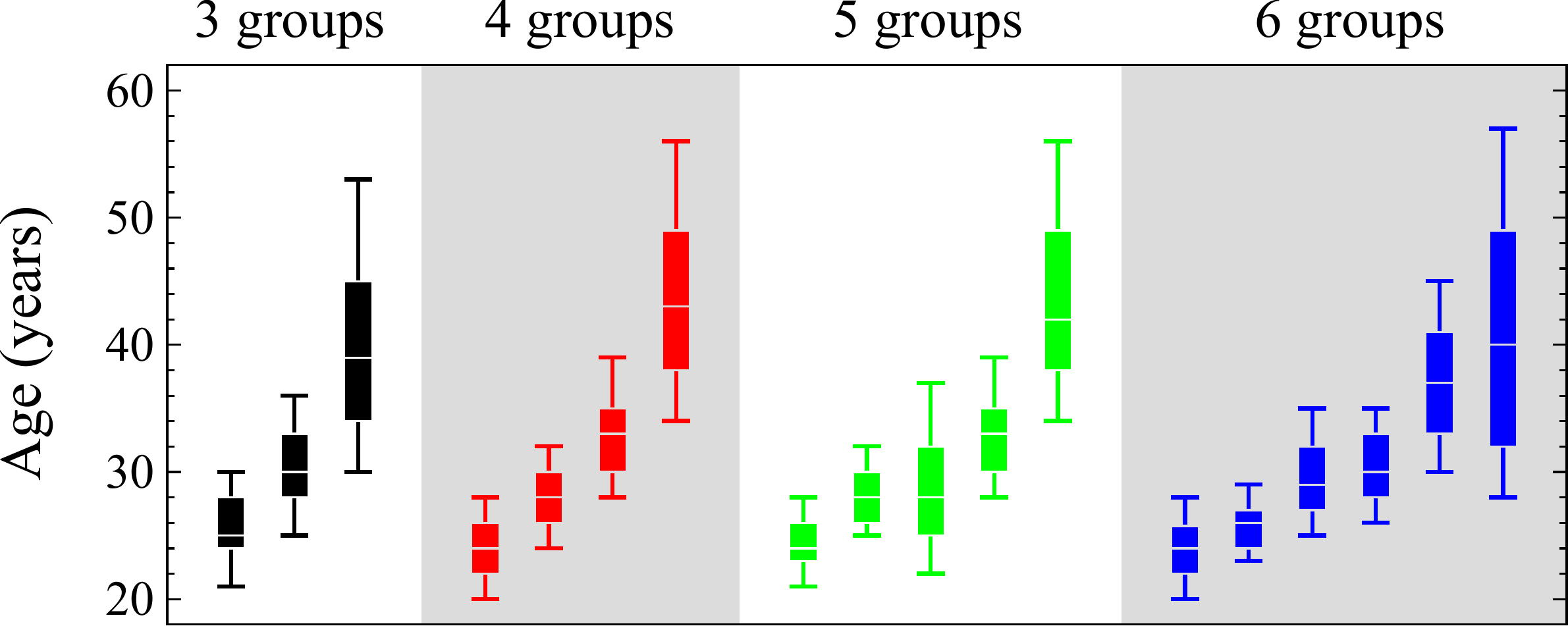}
\caption{Box plots of the age ranges within submarkets for divisions of the New York City user population into three, four, five, and six submarkets.  For clarity, men and women are combined in each submarket in this plot, but a similar pattern is seen when one examines the ages of men and women separately.}
\label{fig:multiway_boxplot}
\end{figure}

\subsection{Number of submarkets}
When applied to the networks of heterosexual dating studied here, the algorithm of the previous section finds clear community structure.  In fact, there are two different types of structure found, one essentially trivial, the other not.  The trivial structure is a division between men and women.  Almost all messages on the web site between heterosexual users looking for romantic relationships are between a man and a woman---well over 99\%.  Very few are between two men or two women.  Our algorithm readily perceives this structure, reliably dividing the network into men and women without the need for us to identify the sexes explicitly.  This ``disassortative'' structure is characterized by a matrix~$\omega_{rs}$ of probabilities that has almost all of its weight off the diagonal (most connections are between different groups) and virtually none on the diagonal (connections between members of the same group).

In addition to this trivial structure, however, there is also the nontrivial group structure that we refer to as submarkets---the tendency of the population to break up into distinct communities of dating with relatively little message traffic between communities.

A practical upshot of this is that if we wish to divide our network into, say, four submarkets, we must actually instruct our algorithm to look for twice this number of communities (i.e.,~eight).  If we do this, then it reliably finds four submarkets, each further divided into men and women.

\begin{figure}
\centering
\includegraphics[width=\columnwidth]{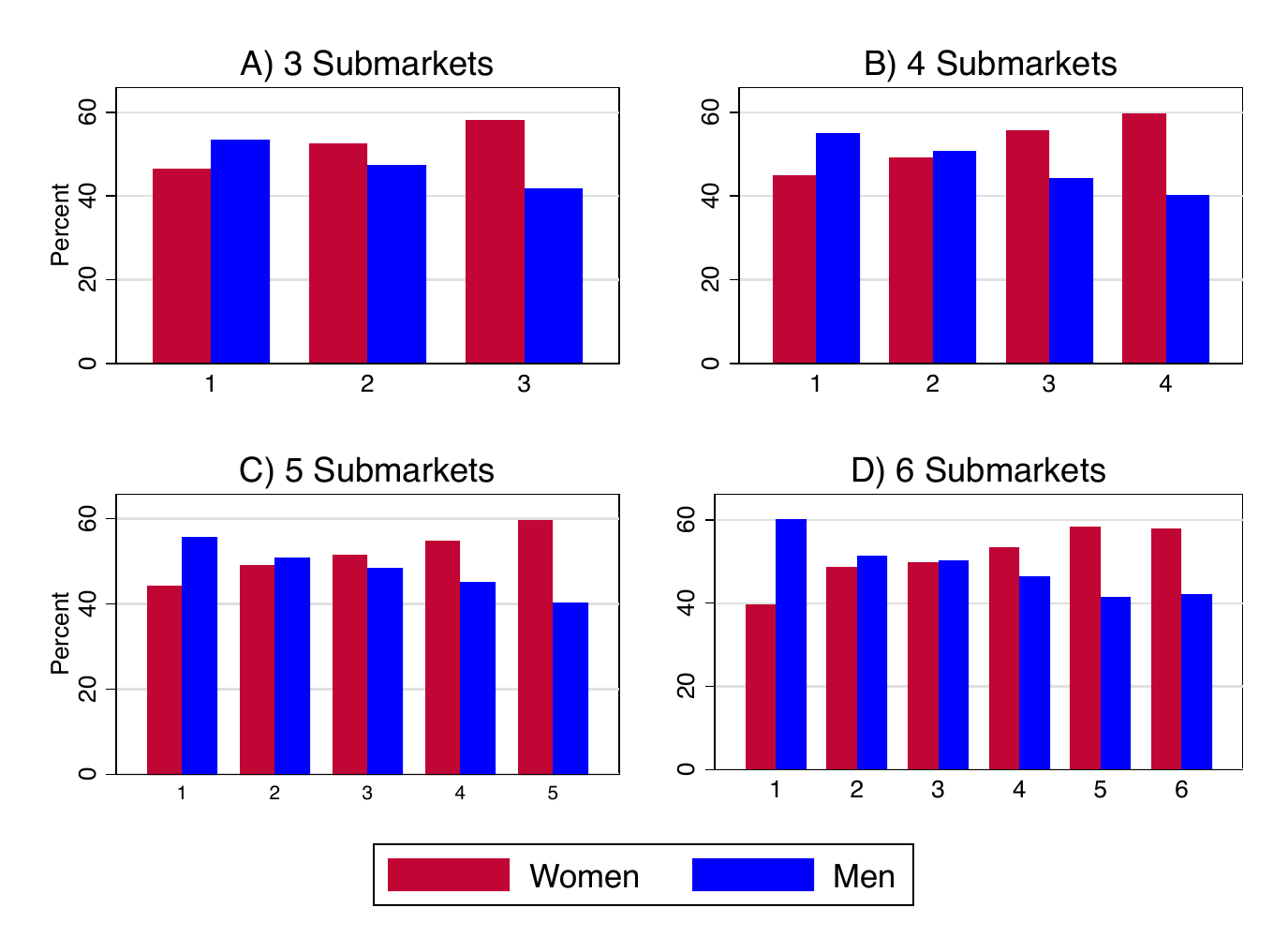}
\caption{Fractions of men and women in each submarket for divisions of the New York City dating market into three, four, five, and six submarkets.  The systematic pattern, seen in Fig.~\ref{fig:submarkets}B, in which the ratio of men to women becomes progressively more female-heavy as we move into the older submarkets, is duplicated in each case here, demonstrating that this is a general behavior, and is not particular to any one choice of the number of submarkets.}
\label{fig:NYC_ratio}
\end{figure}

In the calculations presented in Section~\ref{sec:stratification} we chose to divide each city into four submarkets, but divisions into other numbers of submarkets would also be reasonable.  To explore the effect of varying the number of submarkets we have performed divisions of the networks into various numbers of communities.  Figure~\ref{fig:multiway_boxplot} shows the results of several possible divisions of the New York network.  (Similar patterns are seen in the other three cities.) The panels of the figure show the age distribution (men and women combined) for divisions into three, four, five, and six submarkets (which means six, eight, ten, and twelve communities in total, once the trivial division between men and women is factored in).  As we can see, the primary effect of increasing the number of submarkets is to divide the population into more closely spaced age ranges, so that divisions into larger numbers of groups give a finer, more granular, picture of the market structure but the same overall behavior.  As with all statistical analyses in which data are divided into bins, there is a balance to be struck between larger numbers of bins, which gives finer detail in the analysis, and smaller numbers of bins, which gives better statistics.  Our choice of four submarkets per city gives a good picture of the overall behavior while maintaining sufficient statistical power for accurate analysis of the population within submarkets.

The systematic variation of the ratio of numbers of men and women among submarkets seen in Fig.~\ref{fig:submarkets}B also extends to divisions into other numbers of submarkets, as shown in Fig.~\ref{fig:NYC_ratio}.  As the figure shows, the pattern for the four-way division of Fig.~\ref{fig:submarkets}B, whereby the sex ratio becomes progressively more female-heavy as we move into the older submarkets, is duplicated for divisions into three, five, and six submarkets as well.

\begin{figure}
\centering
\includegraphics[width=\columnwidth]{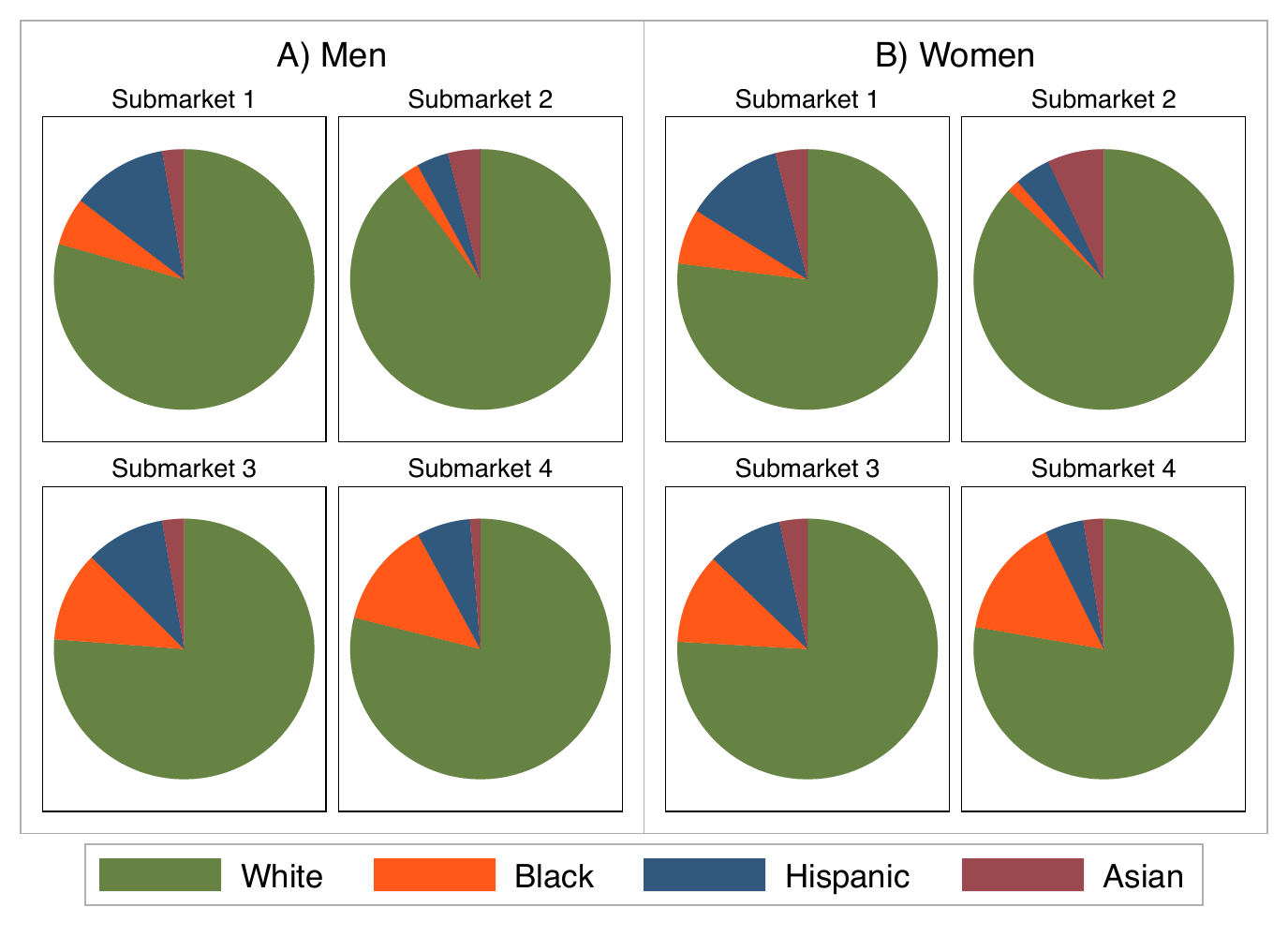}
\caption{Racial composition of submarkets in Chicago. All submarkets are predominantly white, which is consistent with the overall composition of the Chicago market. However, despite the fact that whites are older, on average, than other racial groups, they are disproportionately concentrated in submarket 2. Black users, especially black women, are over-represented in the older submarkets.  Figure~\ref{fig:agediff} suggests one mechanism driving these patterns.}
\label{fig:Chicago_raceComp}
\end{figure}

\section{Additional analyses and results}
\label{app:additional}
In Section~\ref{sec:stratification} we observed that minority women tend to be younger than white women in the same submarket, a trend that is particularly noticeable for black women.  While the pattern holds across all of our four cities, it is most pronounced in Chicago.  Here we provide additional details on the racial composition of Chicago users and insight into processes that give rise to the age differences we observe between white and black women in Chicago. We also examine whether the patterns observed in Chicago hold in New York, the other city with a sizable black population. 

\begin{figure}[b]
\centering
\includegraphics[width=\columnwidth]{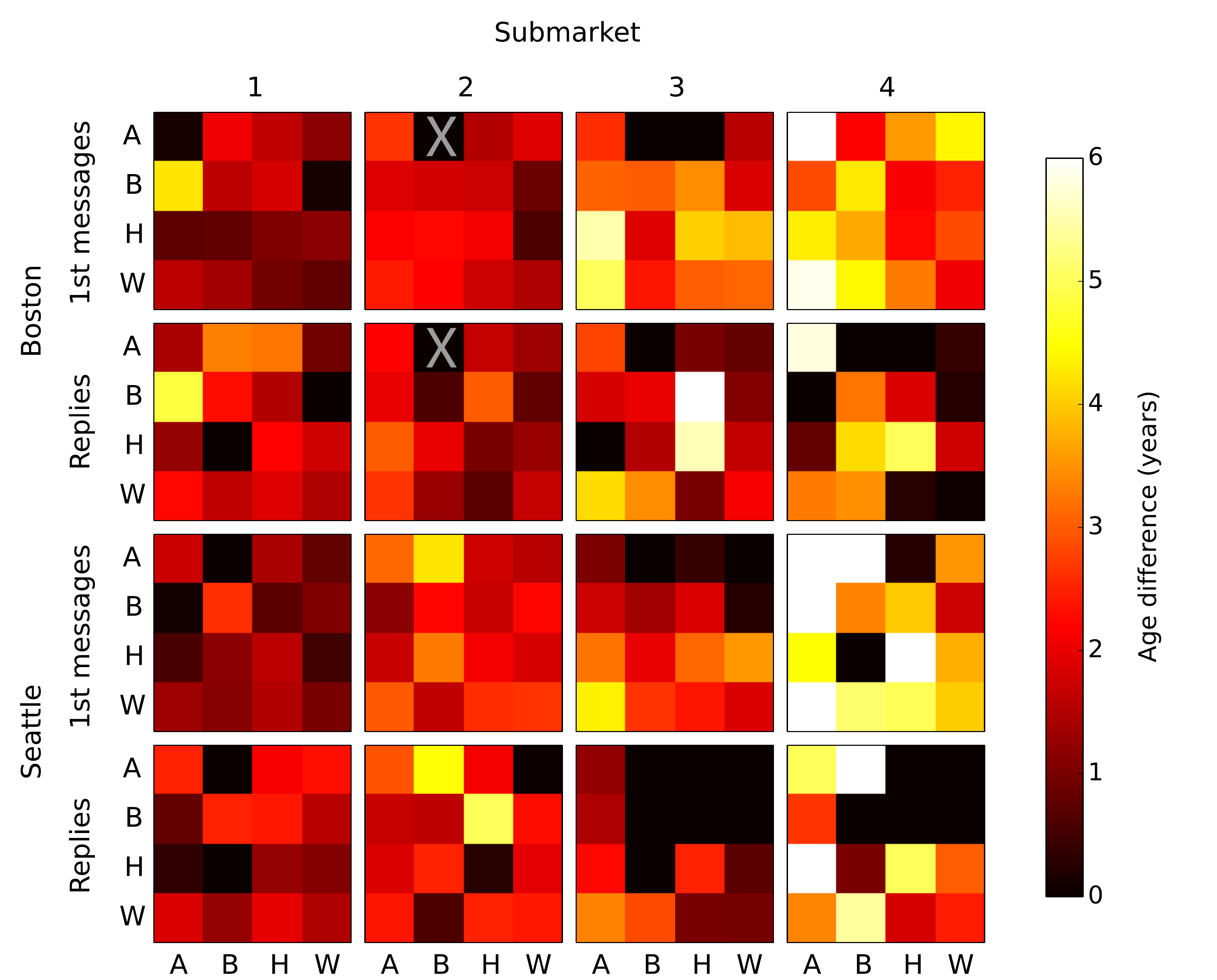}
\caption{Mean difference in years between the age of men of varying races in Seattle and Boston (vertical axis) and the women they message, by race of women and submarket (horizontal axis).  Race is coded as: A = Asian, B = black, H = Hispanic, and W = white.  The first two rows show the average age difference for, respectively, all initial messages sent in Boston and those that received a reply.  The bottom two rows show the same patterns for Seattle. We observe zero instances in Boston where black women receive messages from Asian men in submarket 2, so these cells are marked with an~X.}
\label{fig:BosSea_ageGap}
\end{figure}

Figure~\ref{fig:Chicago_raceComp} shows the mix of ethnicities for men and women in each Chicago submarket.  The predominant group in all submarkets is whites, which reflects the overall composition of the Chicago user base.  There is, however, systematic variation in the relative size of the minority population across submarkets.  Black men and women are more prevalent in the oldest submarkets, which is surprising given that they are slightly younger, on average, than their white counterparts.  One factor driving this is that the black women messaged by both black and white men are, on average, significantly younger than the white women messaged by men in the same submarket, and this phenomenon is most pronounced in the oldest submarkets.  This tends to pull younger women into the older submarkets, and with them the men that they exchanges messages with.  This helps explain not only why there is a surplus of black women in the oldest submarket, but also why these women are significantly younger, on average, than white women in the same submarket.

Figure~\ref{fig:BosSea_ageGap} extends our analysis of age differences in messaging by submarket and race (Fig.~\ref{fig:agediff}) to Boston and Seattle.  The pattern is similar overall to that for New York and Chicago: age differences tend to be larger for first messages than for replies, and also larger in older submarkets.  In submarket~4, for example, white men initiate contact with Asian women who are around 6 years younger than themselves on average, but receive replies from women who are only around 3.5 years younger.  Also in line with the patterns for New York and Chicago, we see that within a given submarket non-white women tend to receive messages from older men than do white women; this is especially true in submarket~4. 

There are, however, also some striking differences between the results for Seattle and Boston and those for New York and Chicago.  In Boston and Seattle, women in submarket~4 (and for Seattle submarket~3 as well) display little tolerance for overtures from much older men.  Note how in these cities women's replies are predominantly to men of similar age to themselves, despite the fact that men are messaging significantly younger women.  Black women in Seattle for example are receiving overtures from black men about 3.5 years older than themselves on average, but reply primarily to men of about their own age.  Notable exceptions to this behavior are messages from Asian men to Asian women, and from Hispanic men to Hispanic women, which appear to receive replies despite large average age differences.


\begin{thebibliography}{10}
\bibitem{note1}
  One can look at the analysis of Fig.~\ref{fig:communities} as providing some justification for previous studies that focused on demographics within geographically localized communities.  While some mate pursuit does take place at longer distances, behaviors are substantially curtailed by geography and demographic patterns of interest can be seen most clearly in smaller local markets.  Most prior studies do have the drawback of considering only the population as whole, however, and not specifically the population of mate seekers, and of lacking, in most cases, direct data on dating and courtship interactions.
\bibitem{note2}
  There are a small number of ZIP codes on the map that appear to have been assigned to the ``wrong'' community, such as the portion of eastern Arizona that is placed in a community together with Ohio and Indiana.  Despite appearances, however, these assignments are not a data error or a fault in the algorithm.  They are a real phenomenon.  Each of them represents an area that had an usual amount of communication with another, far-away locale.  In the eastern Arizona ZIP code, for example, 13\% of conversations were with users in Cleveland, OH, a far larger percentage than with any other area, so the community detection algorithm correctly places this ZIP code in the same community as Cleveland.  Such anomalies will inevitably occur occasionally, usually in sparsely populated areas where the messages sent by even one uncommonly active user can bias the statistics substantially.
\bibitem{note3}
  Black women are over-represented in the oldest submarkets in Chicago and Boston, relative to their fractions in the population, but not in New York or Seattle.  Chicago and Boston are also the only cities where black men tend to be younger than non-black men within the oldest submarket.  This may be because they are pursuing black women, who, within a given submarket, tend to be younger than their non-black counterparts.
\end{thebibliography}
\begin{thebibliography}{10}
\bibitem{note4}
  A CBSA is defined to be an urban center of at least $10\,000$ people plus adjacent areas that are socioeconomically tied to the urban center by commuting.
\end{thebibliography}
\end{document}